\documentclass[sigconf,10pt]{acmart}

\usepackage{graphicx}
\usepackage{textcomp}
\usepackage{xcolor}
\usepackage{subcaption}
\usepackage{booktabs}
\usepackage{multirow}

\usepackage[capitalize]{cleveref}
\usepackage{caption}
\usepackage{subcaption}
\usepackage{float}
\usepackage[htt]{hyphenat} % for text wrapping in \texttt sections
\usepackage{array}
\usepackage{arydshln}

\usepackage{svg}
\usepackage{algorithm}
\usepackage{algpseudocode}
\usepackage{pifont}% http://ctan.org/pkg/pifont
\usepackage{lipsum}  
\usepackage{diagbox}
\usepackage[nolist]{acronym}
\usepackage{listings}
\usepackage[export]{adjustbox}

\Crefname{section}{Section}{Sections}
\crefname{section}{Sec.}{Secs.}
\Crefname{table}{Table}{Tables}
\crefname{table}{Tab.}{Tabs.}
\Crefname{figure}{Figure}{Figures}
\crefname{figure}{Fig.}{Figs.}
\crefname{lstlisting}{List.}{Lists.}

\begin{acronym}
    \acro{SfM}{Structure-from-Motion}
    \acro{NVS}{novel view synthesis}
    \acro{GS}{Gaussian Splatting}
    \acro{GSVC}{Gaussian Splat Video Compression}
    \acro{3DGS}{3D Gaussian Splatting}
    \acro{4DGS}{4D Gaussian Splatting}
    \acro{MV-Frame}{Multiview Frame}
    \acro{VDA}{Video Depth Anything}
    \acro{LOD}{Levels of Detail}
    \acro{MoQ}{Media over QUIC}
    \acro{MoQT}{Media over QUIC Transport}
    \acro{VOD}{Video on Demand}
    \acro{SVC}{Scalable Video Coding}
    \acro{QoE}{Quality of Experience}
    \acro{GOP}{Group of Pictures}
    \acro{DASH}{Dynamic Adaptive Streaming over HTTP}
    \acro{HAS}{HTTP Adaptive Streaming}
    \acro{ABR}{Adaptive Bitrate}
    \acro{VR}{virtual reality}
    \acro{HMD}{Head-Mounted Display}
    \acro{NVC}{Neural Video Coding}
    \acro{NIC}{Neural Image Coding}
    \acro{QP}{Quantization Parameter}
    \acro{ML}{machine learning}
    \acro{NE}{neural-enhanced}
    \acro{NH}{neural-hybrid}
    \acro{MOS}{Mean Opinion Score}
    \acro{PSE}{photosensitive epilepsy}
    \acro{FDE}{Filter Decision Engine}
    \acro{WM}{Workflow Manager}
    \acro{MDC}{Multiple Description Coding}
    \acro{HLS}{HTTP Live Streaming}
    \acro{FPG}{Frames Per Group}
\end{acronym}

\setcopyright{acmlicensed}
\copyrightyear{2018}
\acmYear{2018}
\acmDOI{XXXXXXX.XXXXXXX}
\acmConference[Conference acronym 'XX]{Make sure to enter the correct
  conference title from your rights confirmation email}{June 03--05,
  2018}{Woodstock, NY}
\acmISBN{978-1-4503-XXXX-X/2018/06}

\begin{document}

\title{Point Cloud Streaming with Latency-Driven Implicit Adaptation using MoQ}

% \author{Anonymous Authors}
% Suggested author list: Andrew Freeman, Michael Rudolph, Tanvir Redoy, Finn Schnier, Harrison Hassler, Samira Afzal, Amr Rizk
\author{Andrew C. Freeman}
\email{andrew_freeman@baylor.edu}
\orcid{0000-0002-7927-8245}
\affiliation{%
  \institution{Baylor University}
  \city{Waco}
  \state{Texas}
  \country{USA}
}

\author{Michael Rudolph}
\email{michael.rudolph@ikt.uni-hannover.de}
\orcid{0000-0001-5566-7783}
\affiliation{%
  \institution{Leibniz University Hannover}
  \city{Hannover}
  \country{Germany}
}

\author{Tanvir Redoy}
\email{tanvir_redoy1@baylor.edu}
\orcid{0009-0000-6078-3463}
\affiliation{%
  \institution{Baylor University}
  \city{Waco}
  \state{Texas}
  \country{USA}
}

\author{Finn Schnier}
\email{finn.schnier@stud.uni-hannover.de}
\affiliation{%
  \institution{Leibniz University Hannover}
  \city{Hannover}
  \country{Germany}
}

\author{Harrison Hassler}
\email{harrison_hassler1@baylor.edu}
\affiliation{%
  \institution{Baylor University}
  \city{Waco}
  \state{Texas}
  \country{USA}
}

\author{Samira Afzal}
\email{samira_afzal@baylor.edu}
\orcid{0000-0003-4779-3936}
\affiliation{%
  \institution{Baylor University}
  \city{Waco}
  \state{Texas}
  \country{USA}
}

\author{Amr Rizk}
\email{amr.rizk@ikt.uni-hannover.de}
\orcid{0000-0002-9385-7729}
\affiliation{%
  \institution{Leibniz University Hannover}
  \city{Hannover}
  \country{Germany}
}

\renewcommand{\shortauthors}{Freeman et al.}

\begin{abstract}
    Point clouds are a promising video representation for virtual and augmented reality. Their high-bitrate, however, has so far limited the practicality of live streaming systems. In this work, we leverage the delivery timeout feature within the Media Over QUIC protocol to perform implicit server-side adaptation based on an application's latency target. Through experimentation with several publisher and network configurations, we demonstrate that our system unlocks a unique trade-off on a per-client basis: applications with lower latency requirements will receive lower-quality video, while applications with more relaxed latency requirements will receive higher-quality video.
    
    % higher throughput and quality of point cloud streams as an application's latency threshold increases.
\end{abstract}




\maketitle

% Figure link (open in draw.io): https://drive.google.com/file/d/1SAmn_aKHwuxjpAzBqEIkt1Cz8GWET21o/view?usp=sharing

%% Figures
% Michael will make two tables: throughput and PCQM
% Look at the CHANGE in pcqm and throughput for each video--don't average all the videos together
% Andrew: note in the text that the videos have wildly different source bitrates, so the representations can have dramatically different sizes. Could put a fixed number of points in each?

\section{Introduction}
As virtual reality (VR) and augmented reality (AR) systems gain traction for consumers, the demand for compelling new multimedia experiences is increasing. Point clouds offer a compelling representation for real-time 3D capture; however, their enormous data rates and encoding complexity make point cloud streaming difficult. 

In this paper, we first examine the existing streaming methods relying on pull-based HTTP protocols and client-side adaptation. We then discuss the in-progress Media Over QUIC (MoQ) protocol specification, and argue that its lower overhead and unique transport semantics offer advantages for point cloud streaming. Finally, we introduce a streaming system with multiple description coding, MoQ transport, and implicit adaptation based on an application's target latency.

\begin{figure*}
    \centering
\includegraphics[trim={0.5cm 0.2cm 0.5cm 0.2cm},clip,width=0.85\linewidth]{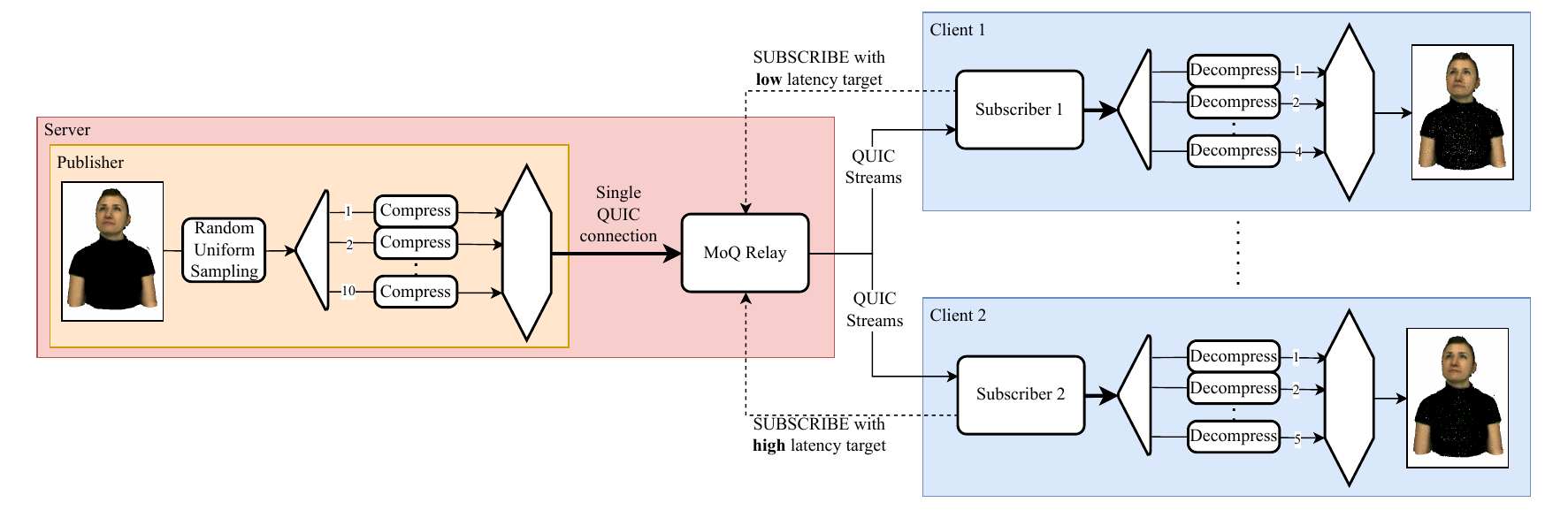}
    \caption{Overall system diagram with an example from our testing. A client that subscribes with a higher latency target will generally receive more representations than a client with a lower latency target, resulting in a higher-quality frame reconstruction. For a given frame, each partition is sent across a unique QUIC stream.}
    \label{fig:overview}
\end{figure*}

Experiments on a real-world dataset demonstrate that our system unlocks a novel trade-off between latency and quality. Real-time applications such as immersive teleconferencing can maintain ultra low latency by receiving a small subset of the original points, while one-to-many live stream viewers with relaxed latency targets can receive higher quality streams. The underlying adaptation is \textit{implicit}, meaning that specific quality selections are not made; rather, we allow the transport layer to indiscriminately drop data as needed based on congestion control and latency targets. Under a reasonable publisher configuration and moderate network congestion, we observe an average 90.8\% increase in throughput and a 0.0024-point improvement in the PCQM metric when increasing the latency allowance from 50 ms to 500 ms. Finally, we demonstrate a proof of concept for content-aware point cloud streaming, using QUIC stream prioritization to increase the throughput of salient points.

\section{Related Work}

\subsection{Point Cloud Streaming}
Inspired by the widespread adoption of \ac{HAS} for video distribution, early work in point cloud streaming proposed extensions of \ac{DASH}~\cite{sodagar2011mpeg} to point cloud content~\cite{hosseini_dynamic_2018}. 
Extensions of this framework allow for viewport-dependent ~\cite{van2019towards} or orientation-dependent~\cite{rudolph2022view, subramanyam2020user} quality adaptation under a bandwidth constraint to increase user quality based on their relative position to the content. 
Using QUIC for transport in HTTP/3 based \ac{DASH} systems for immersive multimedia~\cite{ravuri2023adaptive} allows increased throughput and resilience against packet loss.
However, these systems were designed for on-demand content consumption, but do not meet the low latency requirements for immersive conferencing. In particular, \ac{HAS} relies on \ac{ABR} decision algorithms for explicit quality switching; these algorithms rely on various network performance estimators, which can be inaccurate during high congestion.

Recent work proposed a \ac{MDC} framework for immersive teleconferencing~\cite{deFre2024ScalableMDC}. % \linebreak[4]
Point clouds are divided into  non-overlapping partitions, which are independently coded for transmission.
In contrast to the quality representations offered in \ac{HAS}-based systems~\cite{hosseini_dynamic_2018, van2019towards}, bandwidth adaptation is carried out by varying the number of representations transmitted. 
This has similarities to the scalable coding paradigm introduced for traditional video coding~\cite{schwarz_overview_2007}, which can increase the quality of a base layer representation through a number of enhancement layers. However, scalable coding requires a fixed decoding order, while MDC representations are independently decodable and can be freely recombined. Since point cloud encoding complexity increases with the number of points, MDC also offers a speed advantage for live streaming, as complementary point partitions can be encoded in parallel. 
% Further, point clouds offer a natural fit for deriving non-overlapping partitions. 
Related work uses  Grid-partitioning~\cite{chen2024MDCforPC} to derive a fixed number of representations, which is not perfectly suited for sparse geometries in point clouds.
Random Uniform Sampling of points to receive a sub-group~\cite{deFre2024ScalableMDC} allows the system to freely adjust the number of partitions and the number of points per partition. 

In this work, we explore the MDC paradigm but seek to use delivery timeouts in QUIC to perform implicit, latency-based quality adaptation.

%Using QUIC for transport in a DASH system increases throughput \cite{ravuri2023adaptive,bentaleb_toward_2025}, but quality adaptation remains the responsibility of the client. Adaptation in low-latency HAS systems aims to maintain minimal latency at a best-effort quality. In this work, in contrast, we seek to optimize the received quality for a \textit{wide range of latency targets}.

\subsection{Media Over QUIC}
% In recent years, QUIC has come to constitute much of the everyday Internet traffic that was previously dominated by TCP. The protocol itself is built on UDP, but it is focused on reliable delivery. Compared to TCP, QUIC offers in-built authentication and encryption, enables faster connection initiation, avoids ``retransmission ambiguity,'' and prevents head-of-line blocking \cite{langley_quic_2017}. %Although the protocol resides in user-space, this allows QUIC to be updated more frequently and safely than one implemented in the OS kernel. Indeed, much of QUIC's early development was made possible through large-scale experiments by Google in their Chrome browser \cite{langley_quic_2017}.

% Early on, Google realized the benefits to QUIC-based video transmission in their YouTube product, dramatically reducing the latency for users to begin video playback. Building on these efforts, t
The Internet Engineering Task Force (IETF) has recently been developing a bespoke multimedia streaming protocol built on QUIC (and, by extension, UDP) ~\cite{langley_quic_2017}, termed \ac{MoQ} \cite{gurel_media_2023,nandakumar_media_2025}. Whereas traditional video streaming protocols such as \ac{DASH} and \ac{HLS} rely on HTTP mechanisms for pull-based media delivery, MoQ uses push-based delivery to minimize the end-to-end latency \cite{bentaleb_toward_2025}. The efficacy of push- and UDP-based delivery for low latency is supported by the widespread adoption of WebRTC for teleconferencing applications \cite{sredojev_webrtc_2015}; however, MoQ offers much greater simplicity and flexibility than WebRTC for non-P2P (peer-to-peer) deployments.

The core data abstractions within MoQ are Objects, Groups, Subgroups, and Tracks. For most purposes, an Object will map to a single encoded video frame. Multiple Objects constitute a Group, which is meant to be independently decodable, such as a \ac{GOP} produced through standard video encoding. %Each Group is transmitted across a unique QUIC stream, which helps mitigate head-of-line blocking issues during congestion. 
Groups can be subdivided into Subgroups, which represent related but separately-decodable collections of Objects (e.g., temporal layers in a video). A Track carries all the Groups/Subgroups of a semantic representation, such as a certain-quality video or audio stream. The client can dynamically subscribe or unsubscribe to Tracks as needed to receive the desired content or perform rate adaptation. Each Track publisher will create a new QUIC stream for each Group or Subgroup that it transmits, which helps mitigate head-of-line blocking issues during congestion \cite{nandakumar_media_2025}.

\subsection{Research Gaps}
Prior point cloud streaming solutions largely extend \ac{HAS} semantics, depending on explicit client-side \ac{ABR} logic, leaving the needs of ultra‑low‑latency telepresence unmet. Furthermore, prior efforts have not exploited QUIC stream priorities to preferentially advance salient point subsets when bandwidth is scarce. This paper fills these gaps by implementing and evaluating, to our knowledge, the first MoQ-based pipeline for point cloud streaming. We realize latency-driven implicit adaptation, demonstrate transport-level prioritization for salient content, and provide a controlled comparison against DASH baselines across realistic encoder and grouping configurations. Our system transparently supports both low- and high-latency applications (i.e., both teleconferencing and one-to-many broadcasts) without the overhead of explicit bandwidth estimation or application-specific adaptation algorithms

\section{Method}

% \subsection{Overview}
% We propose a low-latency point cloud streaming system built on \ac{MoQ}. Whereas prior streaming methods perform explicit client-side rate adaptation in the application layer, our approach handles adaptation through the coordination of server-side and QUIC signaling in the transport layer. This allows us to support both low- and high-latency applications without the overhead of explicit bandwidth estimation or application-specific adaptation algorithms. Applications that can tolerate higher latencies will automatically receive more data.

\cref{fig:overview} provides an overview of our system: a video publisher pushes \ac{MDC}-encoded point clouds through a colocated relay to subscriber clients. A client can select a target latency when joining the stream, which implicitly determines the quality of the video they will receive, as the \ac{MoQ} relay attempts to maximize throughput within the selected latency bounds.

\subsection{Point Cloud Sampling}\label{sec:random_sampling}
% This might be better suited to discuss in the experiment setup
Point cloud sampling is the process of selecting a subset of points from a larger point cloud to reduce the data rate, ideally while retaining important visual information. In this early work, we use Random Uniform Sampling as proposed by~\cite{deFre2024ScalableMDC} to derive non-overlapping partitions of the point cloud, using a pre-defined number of partitions. Each partition is separately compressed (in parallel) with the Draco encoder \cite{noauthor_googledraco_2025} and placed into a \ac{MoQ} Object with a corresponding frame ID. 
We disable Draco coordinate quantization to avoid regular gap artifacts between the points when recombining. Although this lowers the compression rate of each partition, our system can react to limited network bandwidth by sending a subset of these partitions. All Objects produced from a given source frame have the same ID, but are published across different \ac{MoQ} Tracks as described below. While we use random sampling as a baseline, the same mechanism can be used to prioritize points corresponding to regions or objects of interest, allowing more perceptually important information to be preserved when only a subset of representations is delivered, as shown below in \cref{sec:salient_sampling}.

\subsection{Session Management}\label{sec:session}
When a publisher begins serving data, it first signals to the \ac{MoQ} relay the number of Tracks it intends to publish. For our system, the Track count matches the number of Draco encoders used, such that each sampled partition is sent across its own Track. These Tracks are assigned QUIC flow control priorities based on their Track IDs, such that lower-ID tracks are more likely to have their data sent during periods of congestion. While it has no bearing on randomly sampled partitions, this prioritization scheme can complement intelligent sampling strategies with highly-salient points allocated to high-priority Tracks (\cref{sec:salient_sampling}).

When a subscriber joins the session, it issues a \texttt{SUBSCRIBE} message for \textit{all} available tracks. Within this message is the \texttt{DELIVERY\_TIMEOUT} parameter. This conveys the receiver's maximum tolerable latency (in ms) for any received Object. %Our key hypothesis was that an increase in delivery timeout will allow more data to propagate to the receiver, especially under periods of high congestion.

% [][TODO: mention that we could just as easily (and probably should) use Subgroups for the proper MoQ semantics]

\subsection{Transport Control}\label{sec:transport}
The relay forwards the received Objects from the publisher to the subscriber. When the publisher first creates a new Object, we insert the current system timestamp in the Object's metadata. Since Objects are served in chunks, we check at each chunk boundary if the delivery timeout has elapsed since the Object was first received at the relay. If the timeout is exceeded, we 
\textit{reset} the underlying QUIC stream associated with that Object's Group. 
% \textit{close} the underlying QUIC stream associated with that Object's Group. This is a slight departure from the \ac{MoQ} draft specification, which dictates that streams be ``reset'' when delivery timeouts are exceeded \cite{nandakumar_media_2025}. A stream reset issues an error code to the stream receiver. When using WebTransport as our QUIC API wrapper in practice, we found that many stream cancellations or resets in rapid succession would cause the entire connection to be closed.
% In our system, however, we do not want the QUIC connection to close during periods of high loss: we simply want to receive as much data as possible within the target latency. When the relay closes a stream due to a delivery timeout, then, we can still 
% The client can then detect this reset at the application layer without a stream error code.
When the receiver attempts to read an Object from a reset stream, it encounters a read error. The receiver interprets this error to mean that any remaining Objects within the Group cannot be delivered. By maintaining a consistent Object ID numbering and Group size, we use this as an explicit signal for Object drops. On the next Group boundary, the subscriber attempts to reopen any reset streams.

% \begin{figure*}
%     \centering
%     \begin{subfigure}{0.3\linewidth}
%         \includegraphics[width=\linewidth]{images/renders/andrew9_rate300_fpg30_encoders10_timeout50_frame250_front.png}
%         \caption{$300$ Mbit/s}
%         \label{fig:render_300}
%     \end{subfigure}
%     \hfill
%     \begin{subfigure}{0.3\linewidth}
%         \includegraphics[width=\linewidth]{images/renders/andrew9_rate600_fpg30_encoders10_timeout50_frame250_front.png}
%         \caption{$600$ Mbit/s}
%         \label{fig:render_600}
%     \end{subfigure}
%     \hfill
%     \begin{subfigure}{0.3\linewidth}
%         \includegraphics[width=\linewidth]{images/renders/andrew9_rate900_fpg30_encoders10_timeout50_frame250_front.png}
%         \caption{$900$ Mbit/s}
%         \label{fig:render_900}
%     \end{subfigure}
%     \caption{Render of frame $250$ of the sequence \textit{andrew9} for increasing available rate, a timeout of $50$ ms and $10$ parallel encoders. Less bandwidth results in less representations available to the decoder.}
%     \label{fig:enter-label}
% \end{figure*}

\subsection{Reconstruction}
When an Object is either received or its Group is dropped for some Track, the receiver notifies a dedicated reconstruction thread. Once this thread has received an Object message for every track, it concatenates the Objects' decoded point arrays. The result is a point cloud with some subset of the original points, which may be saved or displayed as needed. 
% In total, the end-to-end latency, $L$, of a frame is defined by

% $$ L = \max_{i\in I}(E_i) + \min(\max_{i\in I}(D_i), T)$$
% %
% where $I$ is the set of Objects (partitions) for the frame, $E_i$ is the encoding latency for an Object, $D_i$ is the transmission and decoding latency for an Object, and $T$ is the delivery timeout.

% \begin{figure}
%     \centering
%     \begin{subfigure}{\linewidth}
%         \includegraphics[width=\linewidth]{images/throughput/bw_usage_rate300.png}
%         \caption{300 Mbps bandwidth}
%         \label{fig:300Mbps_throughput}
%     \end{subfigure}

%     \vspace{0.5em}
%     \begin{subfigure}{\linewidth}
%         \includegraphics[width=\linewidth]{images/throughput/bw_usage_rate600.png}
%         \caption{600 Mbps bandwidth}
%         \label{fig:600Mbps_throughput}
%     \end{subfigure}
%     \vspace{0.5em}
%     \begin{subfigure}{\linewidth}
%         \includegraphics[width=\linewidth]{images/throughput/bw_usage_rate900.png}
%         \caption{900 Mbps bandwidth}
%         \label{fig:900Mbps_throughput}
%     \end{subfigure}

%     \caption{Average throughput across all videos with different bandwidth limits and MoQ configurations. % A frame is counted as missing if not a single representation is received.
%     }
%     \label{fig:throughputs}
% \end{figure}

\begin{figure*}[t]
    \centering
    \begin{subfigure}{0.23\linewidth}
        \includegraphics[width=\linewidth]{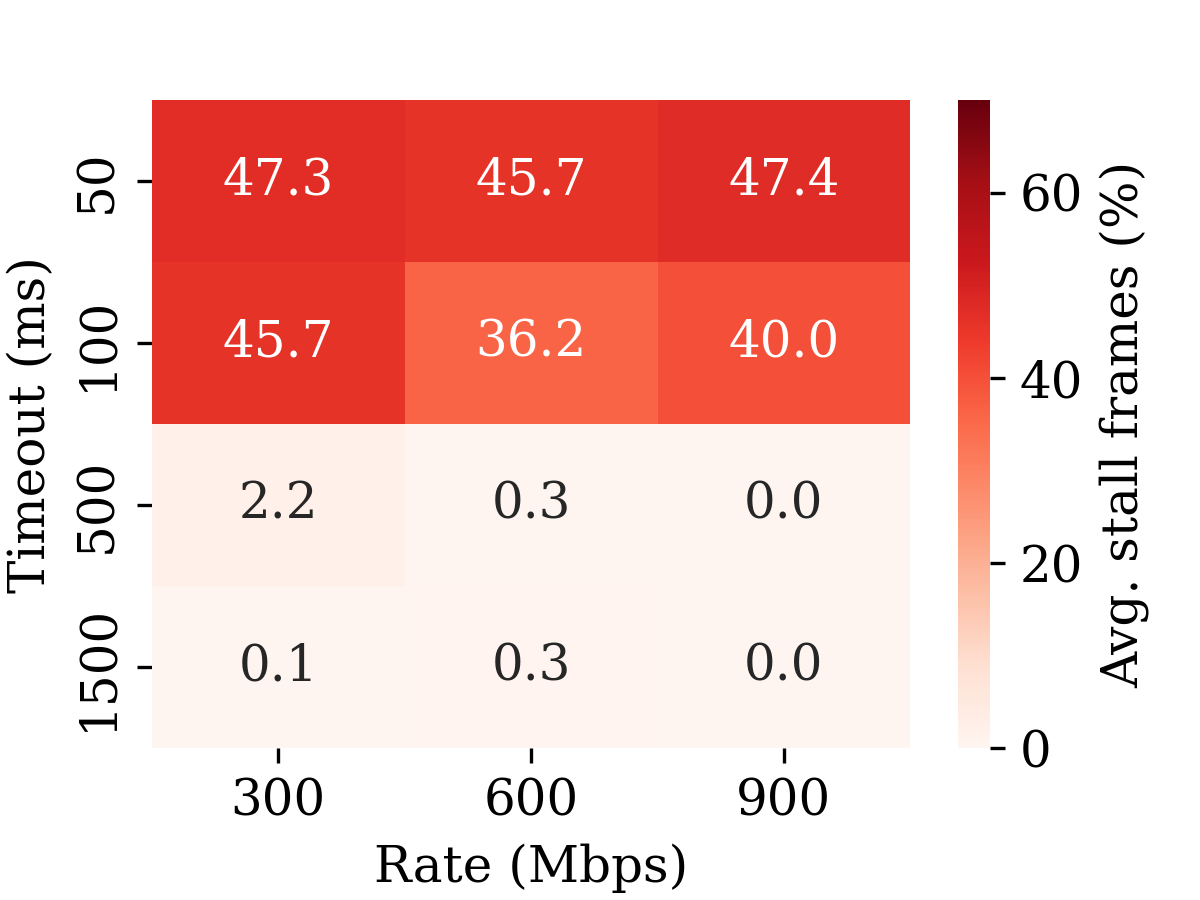}
        \caption{5 Tracks, 5 FPG}
        \label{fig:enc5_gop5}
    \end{subfigure}
    \begin{subfigure}{0.23\linewidth}
        \includegraphics[width=\linewidth]{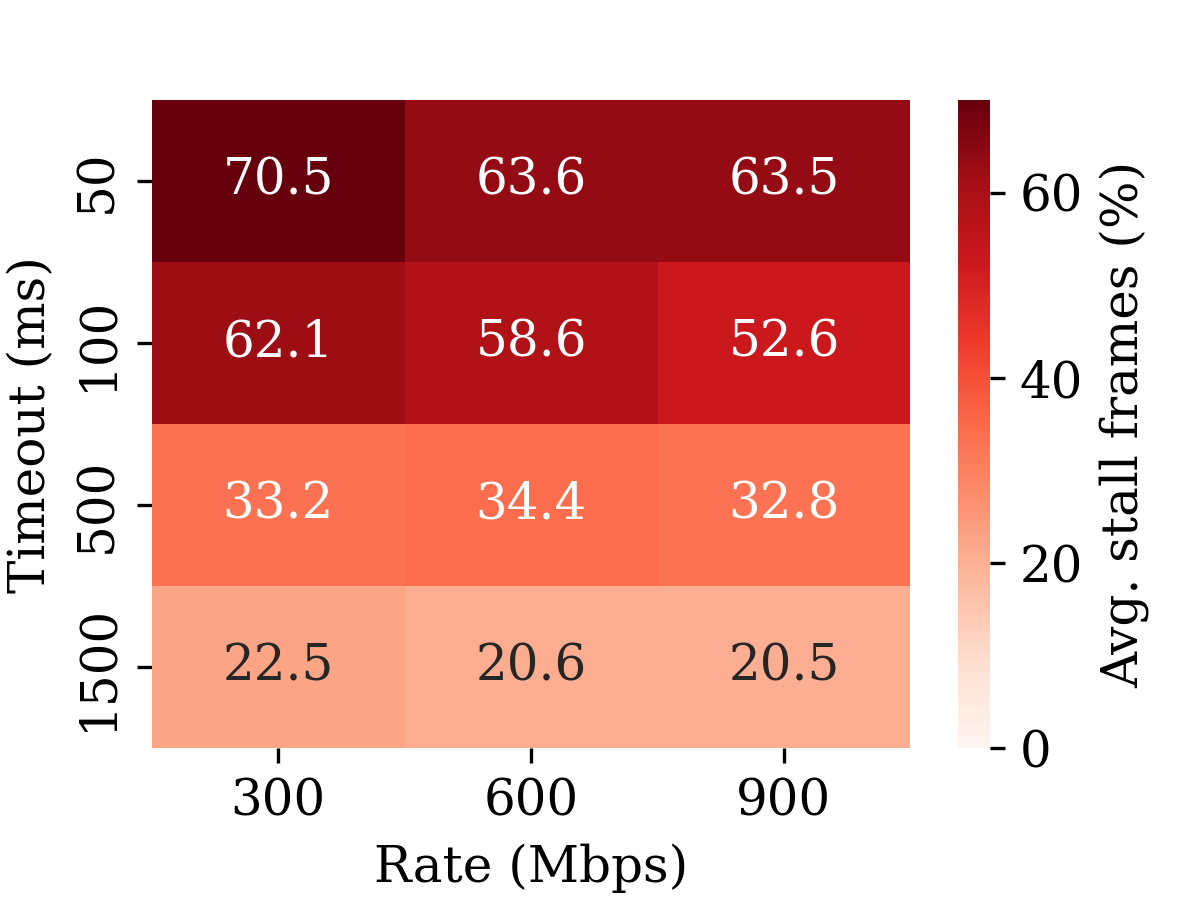}
        \caption{5 Tracks, 30 FPG}
        \label{fig:enc5_gop30}
    \end{subfigure}
    \begin{subfigure}{0.23\linewidth}
        \includegraphics[width=\linewidth]{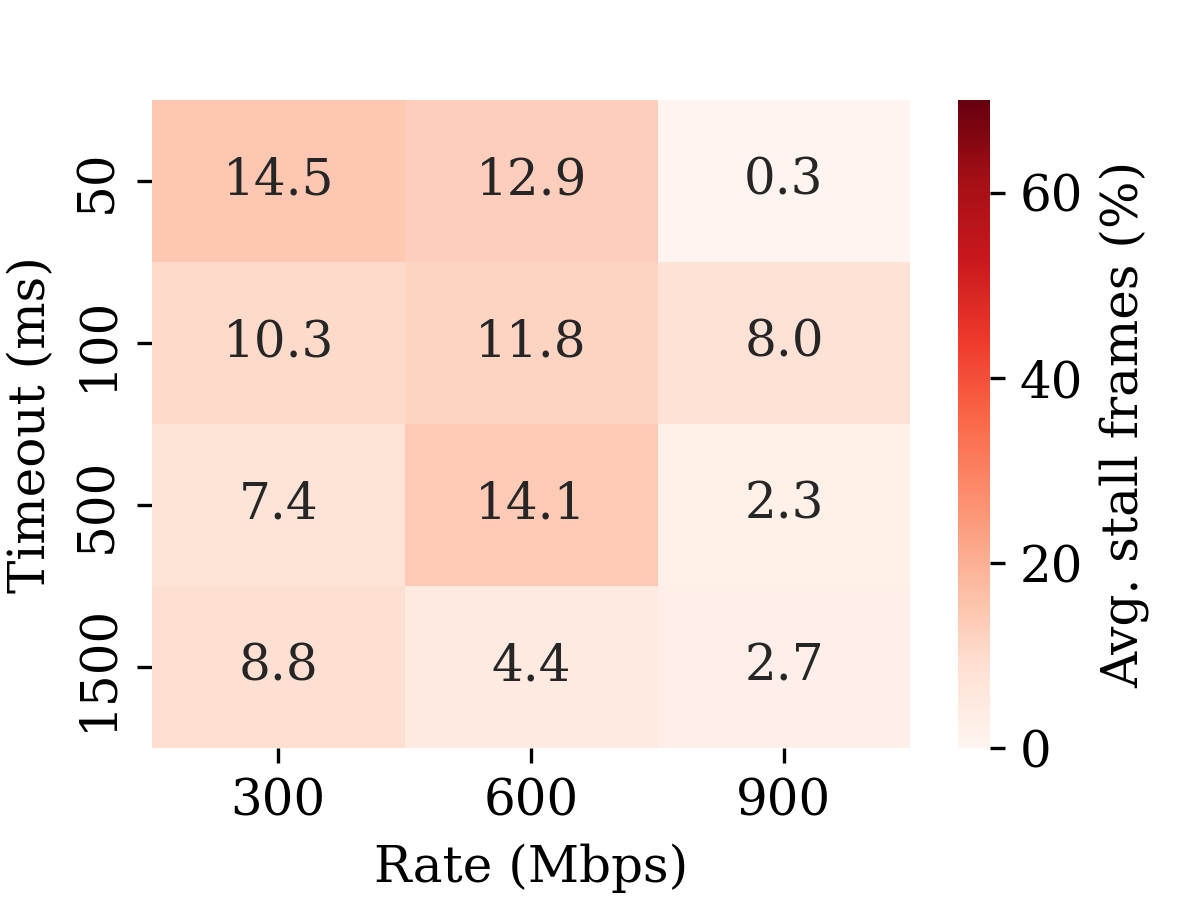}
        \caption{10 Tracks, 5 FPG}
        \label{fig:enc10_gop5}
    \end{subfigure}
    \begin{subfigure}{0.23\linewidth}
        \includegraphics[width=\linewidth]{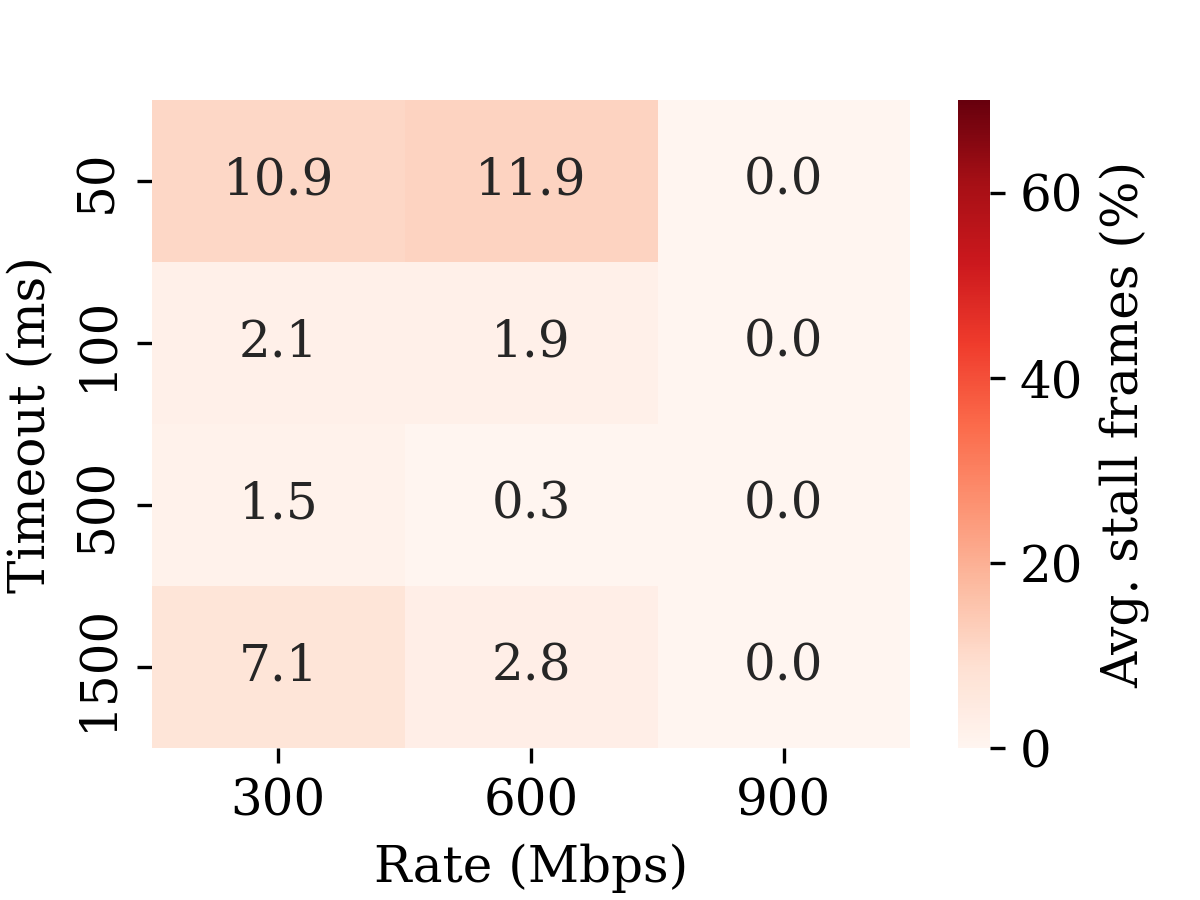}
        \caption{10 Tracks, 30 FPG}
        \label{fig:enc10_gop30}
    \end{subfigure}
    \caption{Stall rate across all test configurations. A frame is counted as stalling if no representation is received. }
    \label{fig:stalls}
\end{figure*}
\begin{table*}
\centering
\setlength{\tabcolsep}{3pt}
\begin{tabular}{cc||cccc|cccc|cccc}
\toprule
\multicolumn{2}{c||}{} & \multicolumn{4}{c|}{300 Mbps} & \multicolumn{4}{c|}{600 Mbps} & \multicolumn{4}{c}{900 Mbps} \\\cline{3-14}
& & \multicolumn{4}{c|}{Timeout (ms)} & \multicolumn{4}{c|}{Timeout (ms)} & \multicolumn{4}{c}{Timeout (ms)} \\
\# Tracks & FPG& 50& $100$ & $500$ & $1500$ & 50& $100$ & $500$ & $1500$ & 50& $100$ & $500$ & $1500$ \\
\hline
5 & 5 & 0.9922 & -0.0001 & +0.0008 & +0.0009 & 0.9941 & -0.0006 & +0.0008 & +0.0011 & 0.9945 & +0.0004 & +0.0038 & +0.0046 \\
5 & 30 & 0.9925 & -0.0001 & +0.0002 & +0.0009 & 0.9946 & -0.0002 & +0.0001 & +0.0007 & 0.9941 & +0.0002 & +0.0020 & +0.0025 \\
10 & 5 & 0.9916 & +0.0004 & +0.0005 & +0.0007 & 0.9953 & -0.0009 & +0.0007 & +0.0009 & 0.9990 & -0.0017 & -0.0007 & -0.0008 \\
10 & 30 & 0.9911 & +0.0005 & +0.0017 & +0.0018 & 0.9940 & +0.0007 & +0.0024 & +0.0014 & 0.9981 & +0.0003 & +0.0009 & +0.0008 \\
\bottomrule
\end{tabular}
\caption{The 50 ms column reports the absolute 1-PCQM, the remaining columns show the average change in 1-PCQM  compared to the baseline 50 ms timeout (positive values indicate quality improvement). Only frames with at least one received representation are included.}
\label{tab:avg_change_pcqm}
\end{table*}

\section{Evaluation}
\subsection{Experimental Setup}

\textit{Point cloud sequences.} We used the Microsoft Voxelized Upper Bodies (MVUB) dataset~\cite{loop2016microsoft} for our evaluation. Each sequence consists of 7-8 seconds of point cloud video recorded at 30 frames per second (FPS) with a spatial resolution of $512^3$ voxels.

\textit{Experimental testbed.} We utilized a testbed consisting of a server with an AMD Ryzen 3700X CPU and a client with an Intel Core Ultra 7 165H CPU. For consistent real-time encoding, we emulated a point cloud production rate at 20 FPS, for streaming durations of approximately 9-11 seconds.

\textit{MoQ setup.} We utilized the Rust-based \texttt{moq} open-source repository as the starting point for our implementation \cite{curley_releases_2025}\footnote{This repository technically implements the simplified MoQ-Lite draft specification \cite{curley_media_2025}, but the mechanisms we employed are held in common between \ac{MoQ} and MoQ-Lite.}.  We implemented the delivery timeout mechanism and its associated control messages as described in \cref{sec:session,sec:transport}. We ran both the publisher and the MoQ relay on our server machine, connected via a network switch on a 1 Gbps link with the client machine.

\textit{Evaluation configurations.} We varied the outgoing bandwidth from the server using \texttt{tc} and \texttt{netem}, examining the rate values 300, 600, and 900 Mbps to induce high, moderate, and low congestion levels, respectively. We ran our publisher with both 5 and 10 encoders and 5 and 30 \ac{FPG}. We initiated our client with delivery timeout values of 50, 100, 500, and 1500 milliseconds. Offline, we computed the Point Cloud Quality Metric (PCQM)~\cite{meynet2020pcqm} scores for the reconstructed images. We report 1 minus the PCQM score (1-PCQM), and assigned a score of 0 if no representations were received for a given frame.
Note that ``acceptable'' 1-PCQM scores should be $>0.98$, with $1.0$ indicating perfect reconstruction~\cite{meynet2020pcqm}.

\textit{DASH comparison.} For a baseline adaptation comparison, we implemented a \ac{DASH}-based streaming system. We leveraged the open-source VV-DASH framework for volumetric streaming \cite{heidarirad_vv-dash_2025}, which includes a \ac{DASH} packager, HTTP server, player, and several standard \ac{ABR} algorithm implementations. Since these algorithms are designed for traditional \ac{ABR} streaming with explicit quality levels (rather than \ac{MDC}), we replicated our \ac{MDC} packaging by encoding a bitrate ladder with progressively increasing numbers of randomly-sampled points. With 5 quality levels, for example, the lowest quality has 20\% of the points, the second-lowest quality has an additional 20\% of the points, etc. Since this approach requires far more encoder computation than true MDC, we could not maintain real-time encoding and packaging at our 20 FPS target. As such, we performed DASH packaging offline, and evaluated it as \ac{VOD} streaming. This yielded the best-case performance for \ac{DASH}, allowing us to isolate the efficacy of explicit \ac{ABR} switching compared to our implicit adaptation. As with \ac{MoQ}, we tested DASH with 5 and 10 quality levels, and segment sizes of 5 and 30 frames. We used the same Draco encoder settings as with the \ac{MoQ} experiments.

\subsection{Results}

% \begin{table*}
%     \centering
%     \input{tables/frames_received}
%     \caption{Percentage of frames received.}

%     \label{tab:my_label}
% \end{table*}

\begin{figure*}[t]
    \centering
    \begin{subfigure}{0.305\linewidth}
        \includegraphics[trim={0.0cm 1.3cm 4.8cm 1.5cm},clip,width=\linewidth]{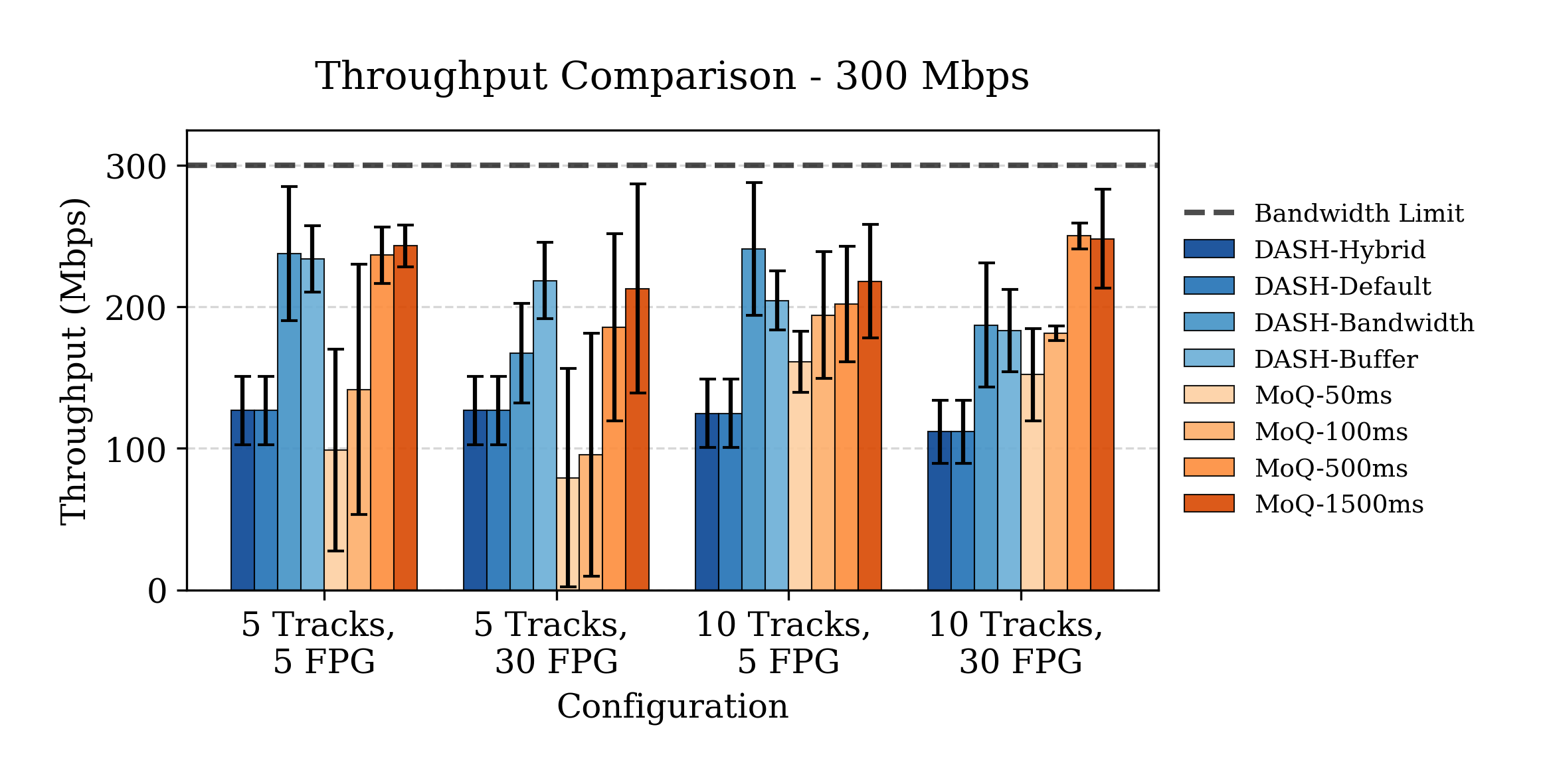}
        \caption{300 Mbps}
        \label{fig:throughcomp1}
    \end{subfigure}
    \begin{subfigure}{0.305\linewidth}
        \includegraphics[trim={0.0cm 1.3cm 4.8cm 1.5cm},clip,width=\linewidth]{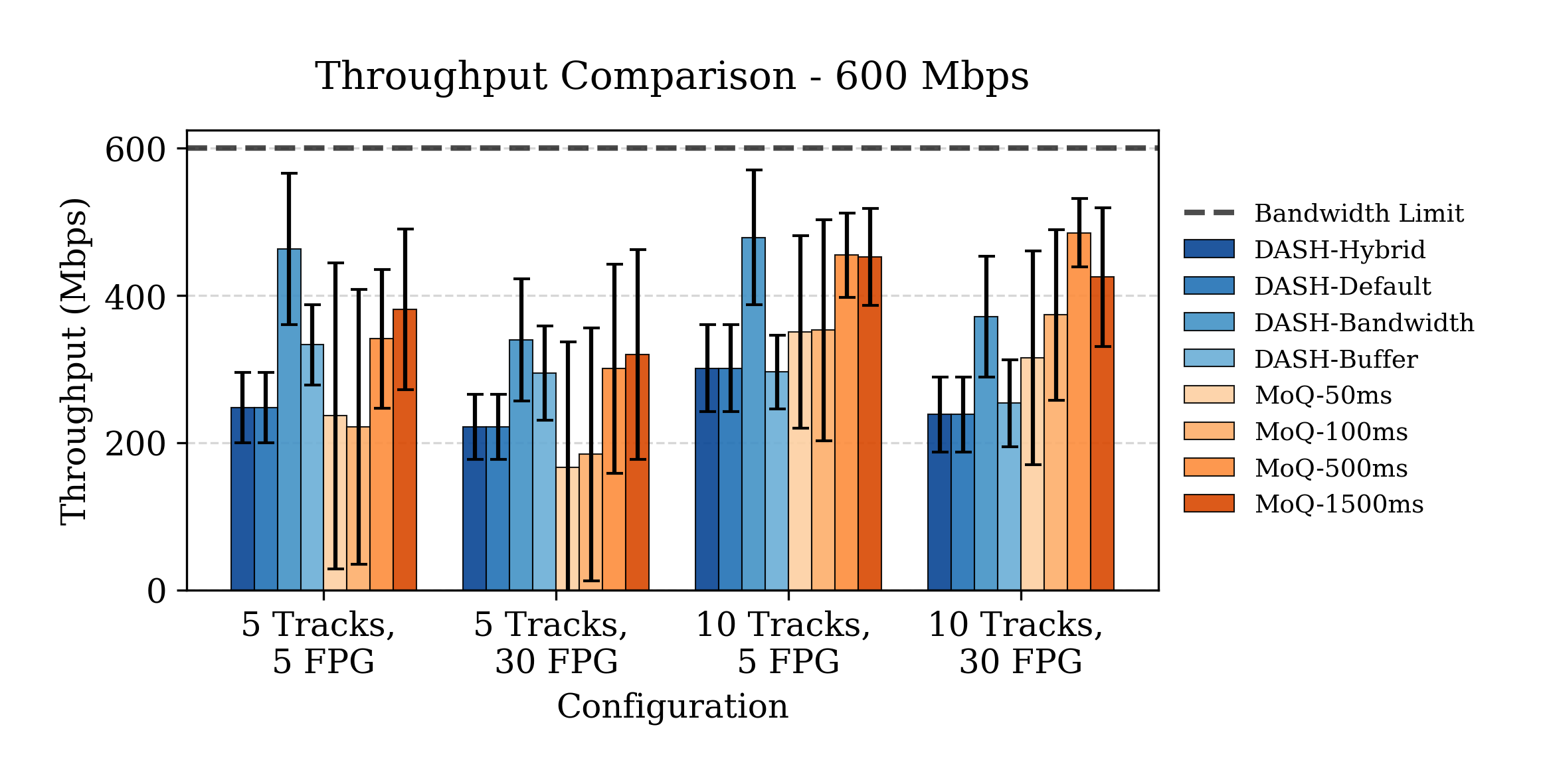}
        \caption{600 Mbps}
        \label{fig:throughcomp2}
    \end{subfigure}
    \begin{subfigure}{0.38\linewidth}
        \includegraphics[trim={0.0cm 1.3cm 0.9cm 1.5cm},clip,width=\linewidth]{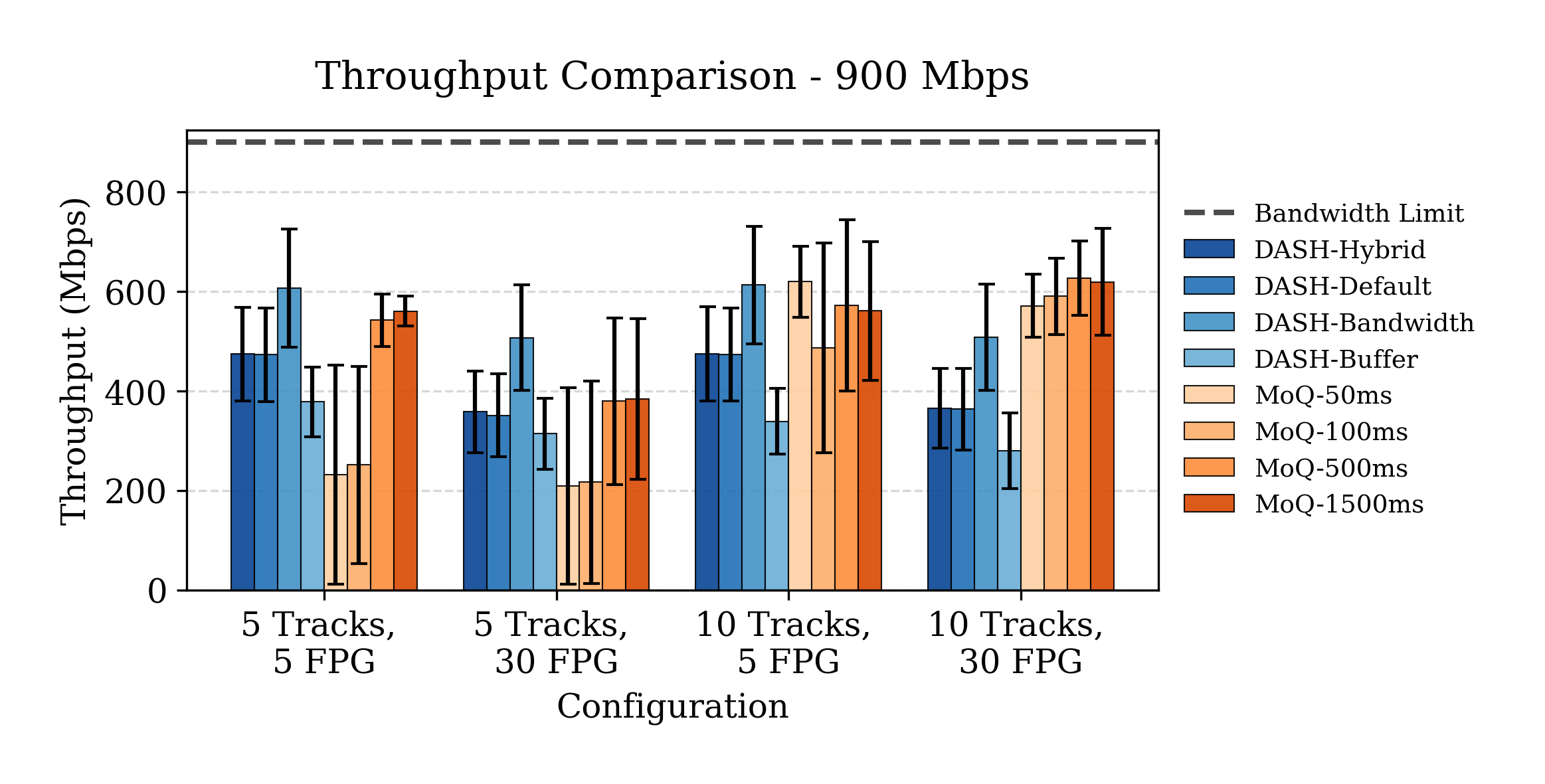}
        \caption{900 Mbps}
        \label{fig:throughcomp3}
    \end{subfigure}
    \caption{Average throughput comparisons between DASH and MoQ for each encoder and packaging configuration. }
    \label{fig:throughcomp}
\end{figure*}

\begin{figure*}[t]
    \centering
    \begin{subfigure}[c]{0.28\linewidth}
        \includegraphics[trim={0.5cm 0.5cm 5.3cm 0.6cm},clip,width=\linewidth]{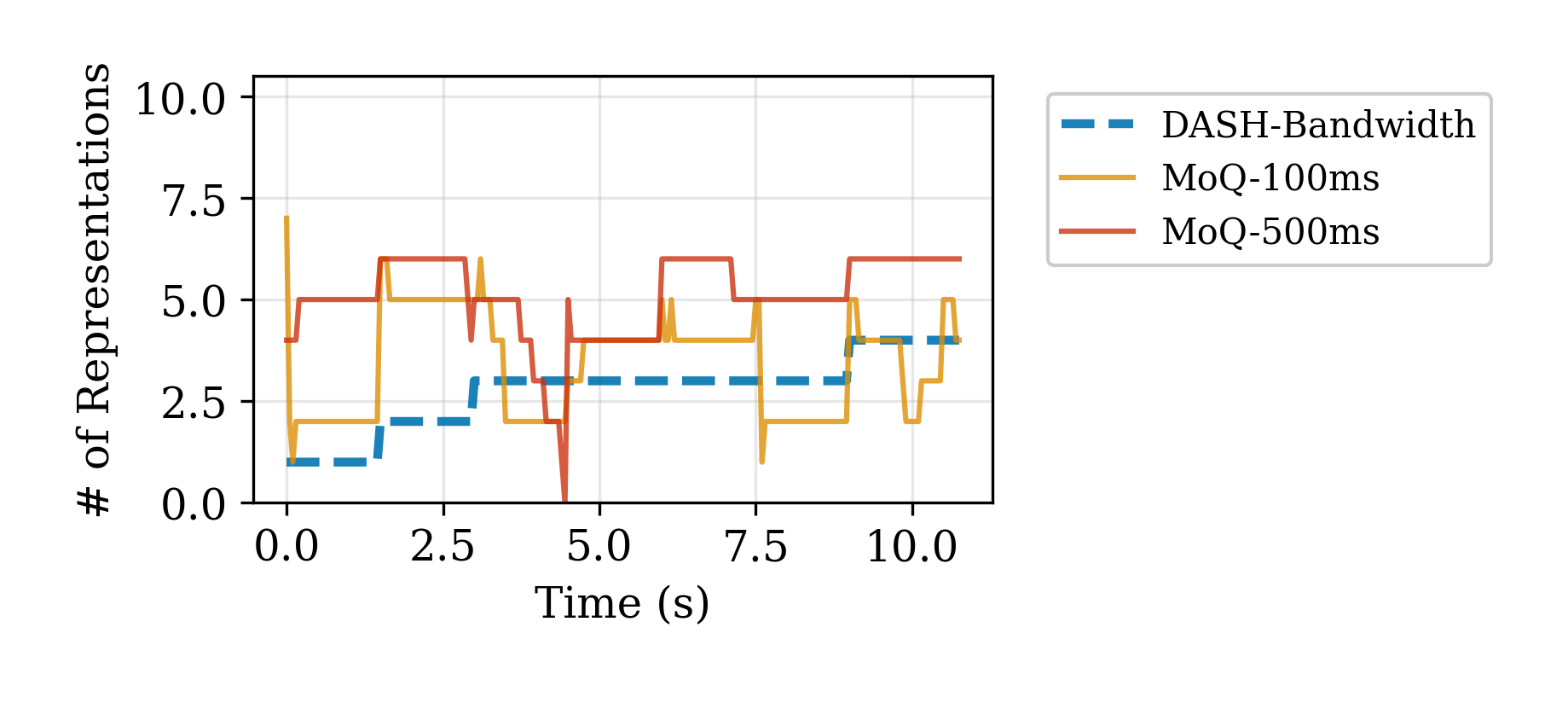}
        \caption{300 Mbps}
        \label{fig:numreps1}
    \end{subfigure}
    \begin{subfigure}[c]{0.28\linewidth}
        \includegraphics[trim={0.5cm 0.5cm 5.3cm 0.6cm},clip,width=\linewidth]{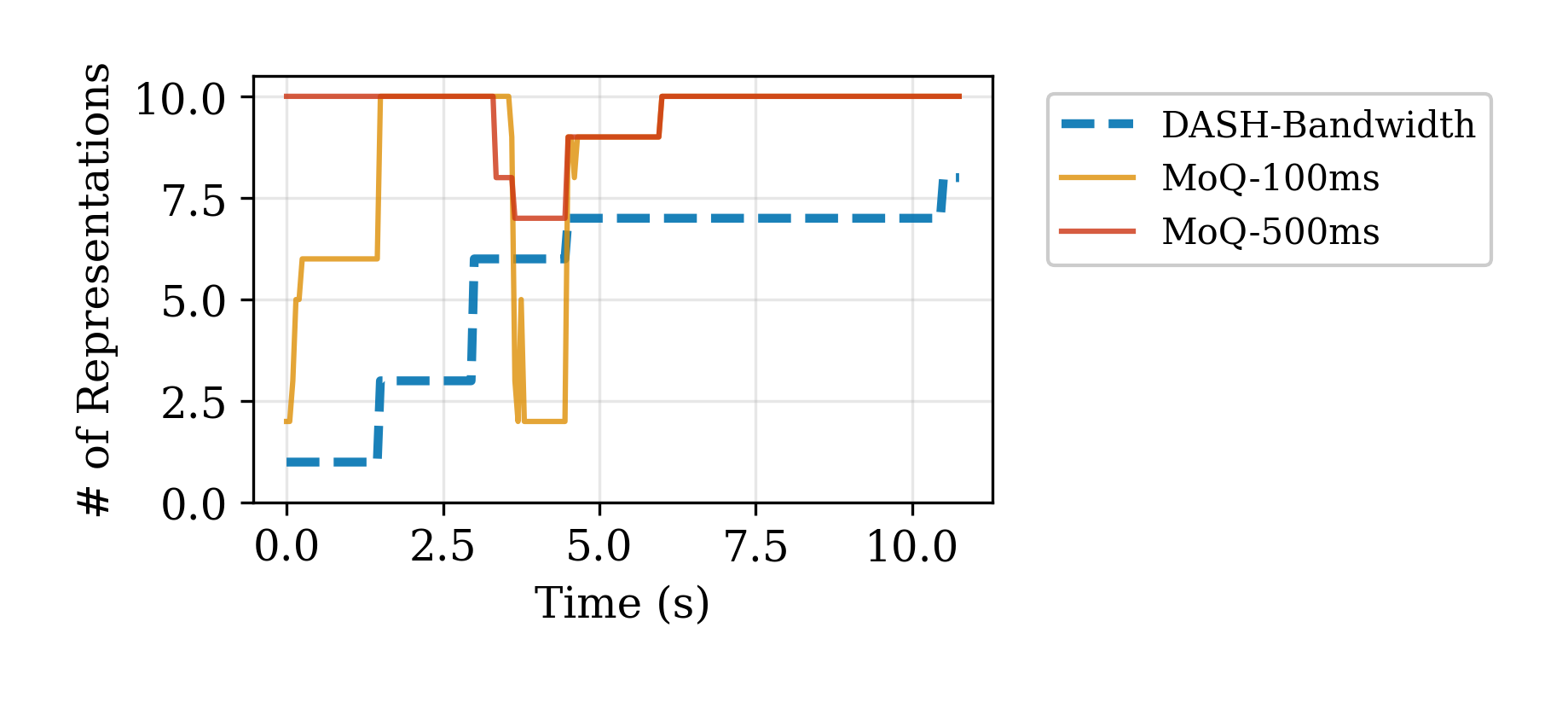}
        \caption{600 Mbps}
    \end{subfigure}
    \begin{subfigure}[c]{0.28\linewidth}
        \includegraphics[trim={0.5cm 0.5cm 5.3cm 0.6cm},clip,width=\linewidth]{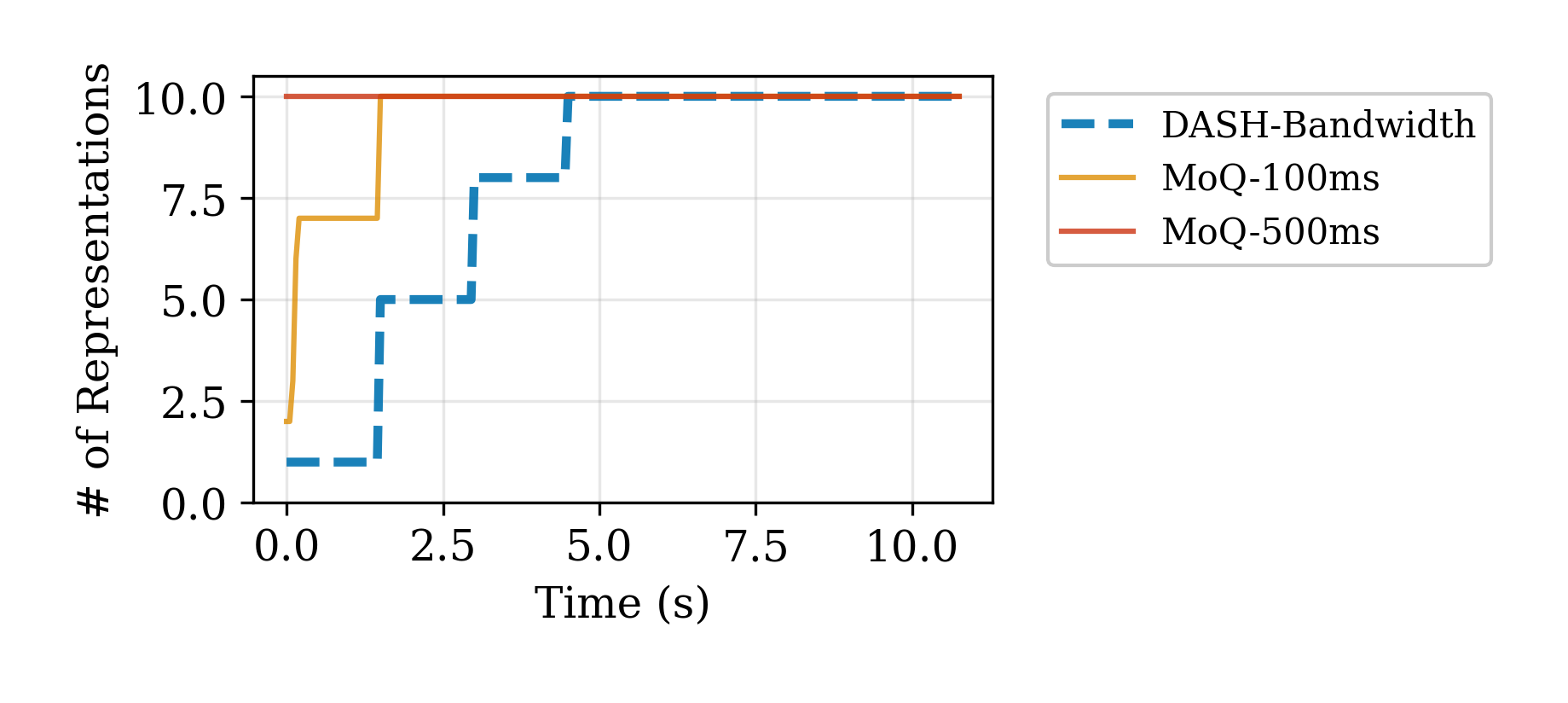}
        \caption{900 Mbps}
    \end{subfigure}
    \centering
    \begin{subfigure}[c]{0.14\linewidth}
        \includegraphics[trim={10.5cm 4.5cm 0.9cm 0.96cm},clip,width=\linewidth]{images/numreps_time/quality_timeseries_ricardo9_rate900_fpg30_enc10.png}
    \end{subfigure}
    % \begin{subfigure}{0.33\linewidth}
    %     \includegraphics[trim={3.0cm 0.5cm 5.9cm 1.5cm},clip,width=\linewidth]{images/numreps_time/quality_timeseries_sarah9_rate300_fpg30_enc10.png}
    % \end{subfigure}
    % \begin{subfigure}{0.33\linewidth}
    %     \includegraphics[trim={3.0cm 0.5cm 5.9cm 1.5cm},clip,width=\linewidth]{images/numreps_time/quality_timeseries_sarah9_rate600_fpg30_enc10.png}
    % \end{subfigure}
    % \begin{subfigure}{0.33\linewidth}
    %     \includegraphics[trim={3.0cm 0.5cm 5.9cm 1.5cm},clip,width=\linewidth]{images/numreps_time/quality_timeseries_sarah9_rate900_fpg30_enc10.png}
    % \end{subfigure}
    \caption{Time series comparison between MoQ and DASH on the \texttt{ricardo9} sequence \cite{loop2016microsoft}, showing the number of received representations at various bandwidth limitations. The configuration with 10 encoders and 30 \ac{FPG} was used. For readability, we show only two MoQ traces, to illustrate the improvement in throughput as the delivery timeout increases.}
    \label{fig:numreps}
\end{figure*}

% [TODO: intro to explain why we looked at the following scenarios]

\subsubsection{Latency–quality trade-off.} We provide the percentage change in average received throughput relative to the 50~ms baseline across our various test conditions in \cref{tab:avg_change_throughput_pct}. Increasing the delivery timeout consistently allows more point cloud partitions to traverse the MoQ relay before QUIC stream resets occur, resulting in higher delivered throughput. This result reflects the system's implicit latency-based adaptation at the transport layer. For example, under congestion at 300~Mbps bandwidth, increasing the timeout from 50~ms to 500~ms yields throughput gains ranging from 27.9\% to 360.2\% depending on the encoding configuration. These throughput gains correspond to improvements in higher reconstruction quality, as shown in \cref{tab:avg_change_pcqm}. The correlation between 1-PCQM scores and received throughput shows an average Pearson coefficient of 0.67, indicating a moderately strong, positive linear relationship.

\begin{table}[h!]
\centering
 \resizebox{\columnwidth}{!}{
\begin{tabular}{cc||ccc|ccc|ccc}
\toprule
\multicolumn{2}{c||}{} & \multicolumn{3}{c|}{300 Mbps} & \multicolumn{3}{c|}{600 Mbps} & \multicolumn{3}{c}{900 Mbps} \\\cline{3-11}
& & \multicolumn{3}{c|}{Timeout (ms)} & \multicolumn{3}{c|}{Timeout (ms)} & \multicolumn{3}{c}{Timeout (ms)} \\
\# Tracks & FPG& $100$ & $500$ & $1500$ & $100$ & $500$ & $1500$ & $100$ & $500$ & $1500$ \\
\hline
5 & 5 & -0.6 & 105.4 & 116.1 & -1.8 & 6889.1 & 6694.4 & 2.8 & 23458.0 & 26226.2 \\
5 & 30 & 55.8 & 360.2 & 712.2 & 65.7 & 258.2 & 644.5 & 65.1 & 936.9 & 3649.1 \\
10 & 5 & 19.7 & 27.9 & 36.7 & 16.5 & 47.4 & 52.0 & -20.5 & -8.3 & -8.1 \\
10 & 30 & 25.6 & 74.1 & 74.8 & 52.7 & 90.8 & 75.2 & 3.2 & 9.8 & 7.9 \\
\bottomrule
\end{tabular}}

\caption{Average change in throughput (\%) compared to the baseline 50 ms timeout.}
\label{tab:avg_change_throughput_pct}
\end{table}

\subsubsection{Stall rate.} In \cref{fig:stalls}, we illustrate the rate of ``stall'' scenarios, which we define as the percentage of frames for which we receive zero representations within the delivery timeout. Using only 5 Tracks results in severe stall rates at short delivery timeouts under low bandwidth, since each Track carries a high data burden. At the same time, when frames are successfully delivered, The 5-Track configurations yield slightly higher per-frame reconstruction quality under low bandwidth (300 Mbps), as reflected by the 1-PCQM scores in \cref{tab:avg_change_pcqm}, which excludes stall frames. In contrast, configurations with 10 Tracks exhibit significantly lower stall rates across all bandwidths and timeout settings, while maintaining high 1-PCQM scores for the non-stall frames. We note that the \ac{QoE} impact of a playback stall is subjective and should be an area of future investigation.

% In \cref{fig:stalls}, we illustrate the rate of ``stall'' scenarios, which we define as frames for which we receive zero representations within the delivery timeout. Although the use of only 5 Tracks yields higher average visual quality under low bandwidth, it also induces severe stall rates at low timeouts. We emphasize that the quality impact of a playback stall is subjective and should be an area of future investigation.

% we clearly see that partitioning the point clouds into only 5 representations yields Objects that are too large to deliver consistently with ultra low latency. 

Under the 10-Track configuration, we observe an improvement when shifting from 5 FPG to 30 FPG. Specifically, we see fewer stall events, higher throughput, and better visual quality. For example, with 600 Mbps bandwidth, we observe an average 90.8\% increase in throughput (\cref{tab:avg_change_throughput_pct}) and a 0.0024-point improvement in the PCQM metric (\cref{tab:avg_change_pcqm}) when increasing the latency allowance from 50 ms to 500 ms. 
%This result is somewhat surprising, as we expected that the faster recovery with smaller Groups would yield higher throughput under network congestion. In practice,
The larger Group size mitigates QUIC congestion through increased spacing between bursts of data.
% as a Track waits longer to begin sending again once a Group is dropped. That is, we better avoid overwhelming the QUIC congestion controller by sending less frequent bursts of data.

\subsubsection{DASH  comparison.} This section compares our MoQ-based system with the four DASH baseline ABR algorithms provided in VV-DASH \cite{heidarirad_vv-dash_2025}. These include the default ``dash'' algorithm, as well as bandwidth-based, buffer-based, and hybrid schemes, which adapt video quality based on network throughput, buffer occupancy, or a combination of both.

\textit{Throughput comparison.} \cref{fig:throughcomp} compares the average throughput between our \ac{MoQ}-based system and the DASH baseline across our three bandwidth settings and all encoder/packaging configurations. Under high congestion, the Bandwidth and Buffer \ac{ABR} algorithms perform best for \ac{DASH}, exceeded only by our 500 ms and 1500 ms delivery timeouts under \ac{MoQ}. Whereas \ac{DASH} has higher throughput for smaller FPG, since it can switch quality level more frequently, \ac{MoQ} exhibits its best average throughput and variance for the configuration with 10 Tracks and 30 FPG. Under this configuration, \ac{MoQ} has similar throughput (and expected PCQM) to the best \ac{DASH} \ac{ABR}, even when we set an extremely low delivery timeout of 50 ms. Meanwhile, we expect that a similarly-configured LL-DASH~\cite{bentaleb_low_2022} implementation would exhibit worse throughput (and quality) than our experiments show, since it has a similar trade-off between quality and latency.

\textit{Dynamic adaptation.} \cref{fig:numreps} illustrates the change in throughput over time for a representative example using the 10 Track, 30 FPG configuration. For readability, we show the number of received point cloud representations (via Tracks or by the selected \ac{DASH} quality level), the high-performing Bandwidth ABR algorithm, and two \ac{MoQ} delivery timeouts. We observe that when network conditions are unknown at the beginning of the stream session, the explicit client-side ABR switching of DASH has a slow warm-up period, gradually increasing its chosen quality. Meanwhile, our \ac{MoQ} system delegates implicit adaptation to the transport layer, attempting to send as much data as possible, allowing for higher initial throughput. However, this also brings with it higher variance in throughput, sometimes leading to stall frames as in \cref{fig:numreps1}.

\section{Content-Aware Streaming PoC}\label{sec:salient_sampling}
We demonstrate a content-aware streaming proof-of-concept (PoC) that leverages QUIC stream priorities to improve visual quality under our implicit adaptation scheme. Instead of Random Uniform Sampling (\cref{sec:random_sampling}), we  variably sample points based on their membership in detected classes of interest. We trained a PointNet model \cite{charles_pointnet_2017} with the EgoBody dataset \cite{avidan_egobody_2022} to perform binary semantic segmentation of human head and non-head classes. This trained model is applied to the MVUB dataset ~\cite{loop2016microsoft} in real time with a GTX 1080 Ti GPU. We then weight our random sampling such that the \ac{MoQ} Tracks with the highest QUIC priorities contain approximately 50\% ``head'' points and 50\% ``non-head'' points. 

\begin{figure}
    \centering
    \begin{subfigure}{0.49\linewidth}
        \includegraphics[trim={-3cm 0cm 3cm 0cm},clip,width=\linewidth]{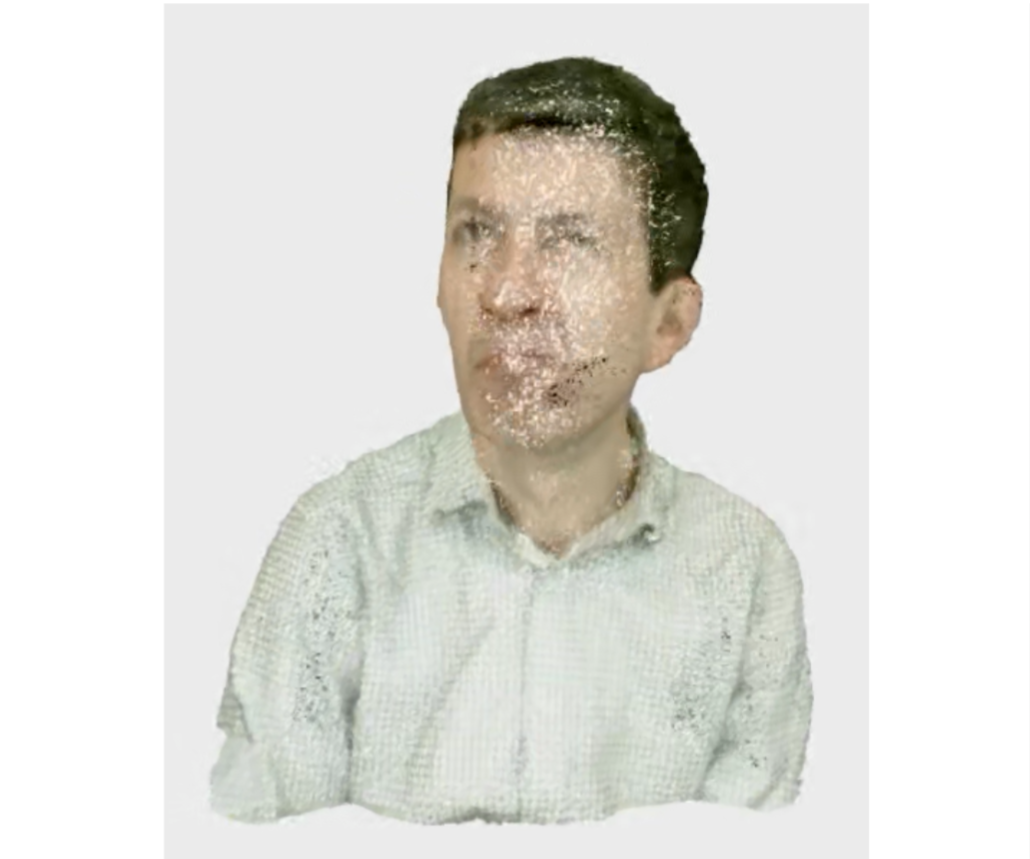}
        \caption{Random, VMAF: 68.9}
    \end{subfigure}
    % \hspace{0.5cm}
    \begin{subfigure}{0.49\linewidth}
        \includegraphics[trim={3cm 0cm -3cm 0cm},clip,width=\linewidth]{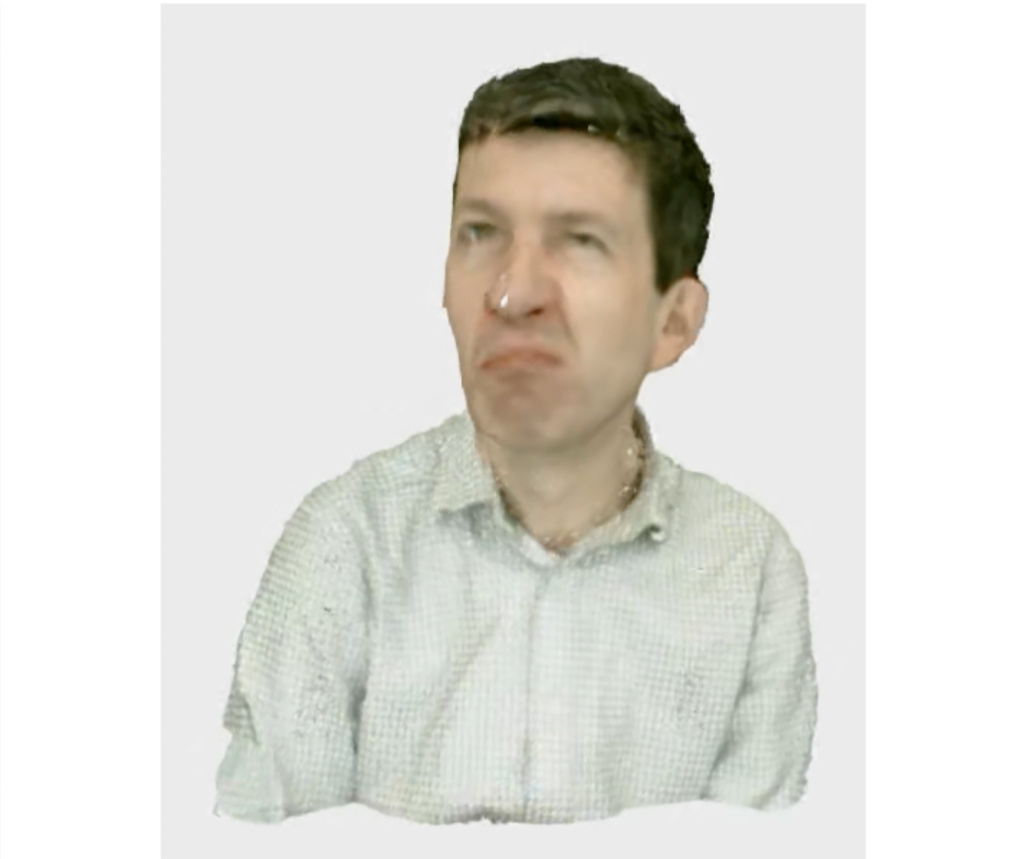}
        \caption{Salient, VMAF: 79.0 }
    \end{subfigure}
    \caption{Impact of content-aware sampling on the \texttt{andrew9} sequence \cite{loop2016microsoft}, with higher transport priority given to points detected in the head region. Configuration: 10 Tracks, 30 FPG, 600 Mbps bandwidth, 100 ms timeout.}
    \label{fig:semanticsegmentation}
\end{figure}

Subject to reasonable segmentation accuracy, we find that our \ac{MoQ} scheme implicitly emphasizes the transmission of the salient ``head'' points during periods of network congestion. We provide a visual example in \cref{fig:semanticsegmentation}, with 2D-rendered frames. We can see that the content-aware sampling yields far fewer holes in the facial region, which is crucial for subjective quality. This yields a 10.1 improvement in average VMAF score under moderate network congestion\footnote{We provide an anonymous video example for the reviewers here: \href{https://zenodo.org/records/18203126}{https://zenodo.org/records/18203126}}. This result demonstrates a unique use case to combine \ac{MoQ} delivery timeouts with deliberate QUIC stream prioritization for content-aware streaming, which may improve \ac{QoE} for immersive teleconferencing applications.

\section{Conclusion and Future Work}
% Conclusion

% \begin{figure}
%     \centering
%     \begin{subfigure}{\linewidth}
%         \includegraphics[width=\linewidth]{images/pcqm/pcqm_rate300.png}
%         \caption{300 Mbps bandwidth}
%         \label{fig:300Mbps_throughput}
%     \end{subfigure}

%     \vspace{0.5em}
%     \begin{subfigure}{\linewidth}
%         \includegraphics[width=\linewidth]{images/pcqm/pcqm_rate600.png}
%         \caption{600 Mbps bandwidth}
%         \label{fig:600Mbps_throughput}
%     \end{subfigure}
%     \vspace{0.5em}
%     \begin{subfigure}{\linewidth}
%         \includegraphics[width=\linewidth]{images/pcqm/pcqm_rate900.png}
%         \caption{900 Mbps bandwidth}
%         \label{fig:900Mbps_throughput}
%     \end{subfigure}

%     \caption{PCQM}
% \end{figure}

We present a point cloud streaming system with implicit server-side adaptation through the use of \ac{MDC} and \ac{MoQ}. The system dynamically selects how many point cloud sample representations to send based on the receiver's target latency, thereby unlocking a novel trade-off between latency and visual quality. We illustrated that applications with stringent low-latency requirements, such as immersive teleconferencing, can maintain minimal delays by receiving a smaller subset of point cloud representations. Conversely, applications with more relaxed latency targets, like one-to-many live streaming, can benefit from higher-quality streams as more data propagates to the receiver within the allowable delay. This implicit adaptation, integrated with our point cloud sampling and QUIC-based transport, enhances the viability of diverse VR/AR applications. A tight coupling of point cloud sampling, compression, and streaming with \ac{MoQ} can unlock more adaptive and high-quality experiences in extended reality systems.

% Future Work
    While Random Uniform Sampling offers the low latency necessary for immersive conferencing, we demonstrated that content-aware partitioning schemes are a promising direction to improve \ac{QoE}. Further work should explore more advanced models with additional content classes.
    The application of scalable coding strategies is an alternative future direction, but requires future exploration on the robustness of MoQ/QUIC prioritization mechanisms.
% This can be elegantly paired with prioritization mechanisms in MoQ, allowing us to give more precedence to representations deemed more important during transmission. For example, human faces should have a higher priority than their clothing for teleconferencing applications.

% Lossy Draco settings and additional point cloud encoders should be further explored in the context of MDC. Advanced sampling strategies may help mitigate artifacts from partial reconstructions. Results may be more consistent if there is an equal number of points per representation; however, this may adversely impact the reconstruction quality for high bitrate videos. 

% Furthermore, the delivery timeout concept can be extended to apply to the encoders themselves, to free up server-side compute resources and mitigate network congestion. The underlying QUIC congestion control algorithms can themselves be explored to improve the implementation of MoQ delivery timeouts: it would be preferable to avoid sending any chunks of an Object if it will ultimately fail to meet the timeout.

Finally, there is significant room to explore the impact of implicit adaptation on the user's \ac{QoE}. Our PCQM measurements indicate a predictable latency-distortion trade-off for individual frames, but we do not yet quantify the impact of potentially drastic changes in the received quality or stall scenarios when no representations are received for a given frame. Subjective user studies will be necessary to evaluate and mitigate these factors. From there, we can better determine the optimal encoder and sampling configurations to maximize \ac{QoE}.

% [Andrew: another idea, but probably not worth mentioning here, is that we don't have to make the Groups aligned to the same start indices. Even we use an encoder with inter-prediction, we could handle it on the receiver the exact same way. This could help reduce the number of "stall" scenarios, assuming that those have a negative impact on viewer experience.

% \section*{Acknowledgment}

%\section*{References}

% \section{Meeting Notes:}
% \subsection{May 7:}
% Next steps:
% Michael
% \begin{itemize}
%     \item Get Draco encoder running
%     \item Create system to read from file for testing
%     \item Apply uniform sampling
%     \item Get encoders running
%     \item Talk about how to integrate with MoQ
% \end{itemize}

% Andrew
% \begin{itemize}
%     \item Once that's done, create C++ bindings or UNIX socket
%     \item Packaging in MoQ
% \end{itemize}

% What to evaluate
% \begin{itemize}
%     \item Latency vs. quality in different bandwidth scenarios (and what the number of original points is)
%     \item Visual quality metrics on point clouds directly
%     \item VMAF on rendered 2D videos with fixed positions/trajectories
%     \item Just looking for these objective metrics to get worse in a predictable way as the latency and bandwidth decrease (max quality doesn't necessarily need to be good)
%     \item Quality Switches/ Stalls
% \end{itemize}

\bibliographystyle{ACM-Reference-Format}
\bibliography{references,references2,references3}

\end{document}